\documentclass[]{article}
\usepackage{amssymb,latexsym}
\usepackage[superscript]{cite}
\usepackage{amsmath}
\usepackage{amsthm}
\usepackage{graphicx}
\usepackage{epsfig}
\usepackage{caption}
\usepackage{subcaption}
\usepackage{float}
\usepackage{wrapfig}
\usepackage{fancyhdr}
\usepackage{tensor}
\usepackage{textcomp}
\usepackage{multicol}
\usepackage{tikz}
\usepackage{tkz-euclide}
\usetikzlibrary{arrows}
\usepackage{circuitikz}
\usepackage{siunitx}
\usepackage{colortbl}
\usepackage{enumerate}
\usepackage{ifthen}
\usepackage[left=3cm,top=3cm,bottom=3cm,right=3cm,columnsep=.6cm]{geometry}

\newcommand{\bs}{\begin{split}}
\newcommand{\es}{\end{split}}

\theoremstyle{plain}

\theoremstyle{definition}

\usepackage[colorlinks=true,linkcolor=blue,citecolor=blue,urlcolor=blue]{hyperref}

\usepackage{authblk}
\title{Quantitative ultrasound characterization and comparison of healthy and malignant prostate cells}
\author[2]{Penelope Taylor}
\author[2]{Amy Longstreth}
\author[1]{Maria-Teresa Herd}
\affil[1]{Department of Biological and Physical Sciences, Assumption College, Worcester, MA, 01609, USA}
\affil[2]{Physics Department, Mount Holyoke College, South Hadley, MA, 01075, USA}
\date{\today}                     

\begin{document}
\begin{titlepage}
\maketitle
\end{titlepage}

\begin{abstract}
Speed of sound and attenuation as a function  of frequency between 2 and 18 MHz were measured and compared for a cancerous prostate cell line and a healthy prostate cell line. Speed of sound for the cancerous cells line was found to be 1521.4 $\pm$ 0.8~m/s, which was equivalent to the speed of sound for the healthy cell line of 1521.5 $\pm$ 0.6~m/s.  The average attenuation coefficient was  0.091 $\pm$ 0.003 $\frac{\rm{dB}}{\rm{cm-MHz}}$ for the cancerous prostate cell line and 0.057 $\pm$ 0.003 $\frac{\rm{dB}}{\rm{cm-MHz}}$ for the healthy  prostate cell line, showing a higher attenuation for the cancerous cell line.  
\end{abstract}






\section{\label{sec:Intro}Introduction}

Quantitative ultrasound (QUS), which includes the measurements of tissue properties, is gaining importance in medicine as a possible diagnostic and detection tool. While ultrasound (US) has limited image resolution, QUS allows for a greater collection of data in addition to the images formed using B-Mode US. Properties measured using QUS, such as scatterer size, attenuation, and speed of sound have all been shown to differentiate between benign and malignant tumors.  For example, there are significant differences in the sizes of scatterers between a breast-tumor model and non-malignant breast tissue \cite{Oelze2002}$^,$ \cite{Oelze2004} and scatterer sizes in liver hemangiomas are somewhat larger than scatterer sizes in the surrounding liver parenchyma \cite{Liu2007}. Additionally, multiple studies that use attenuation and speed of sound (SOS) as cancer diagnostic tools have been performed.  \cite{Glide2007}$^,$ \cite{Glide2007b}$^,$ \cite{Duric2007}$^,$ \cite{kijima2008}$^,$ \cite{Maeva2009}. Measurement of ultrasonic tissue properties can be an important diagnostic tool for distinguishing between malignant and benign tumors, and may possibly be used for detection of cancer. 

The National Institute of Health reports that for men prostate cancer is the most common cancer and the second most common cause of death due to cancer. \cite{NIH1} The most recent data shows that there are about 186 new cases of prostate cancer per 100,000 men, with 24 deaths per 100,000 men each year. \cite{NIH2} Prostate cancer is difficult to detect and diagnose. Currently prostate-specific antigen (PSA) levels are tested, and biopsies are performed if the PSA levels are high. This is not a very accurate indicator, since two thirds of all biopsies are benign.
Several studies on utilizing ultrasound to diagnose and detect prostate cancer have been done, including using spectrum analysis to distinguish between cancerous and benign prostate tissue \cite{Feleppa96}$^,$ \cite{Feleppa97} and using attenuation and backscatter coefficient measurements to show pathological changes in prostate and testis tissue\cite{Gartner98}.  Although one study measuring SOS, attenuation, and backscatter coefficients failed to distinguish between benign and malignant tissue, the study did show differences in the echogenicity of cancerous tissue \cite{Russell98}. Notwithstanding mixed results, each study suggests potential for using QUS to diagnose and detect prostate cancer. Further studies on utilizing QUS for distinguishing between benign and malignant prostate tissue would contribute to this base of knowledge, and clarify the differences found in these studies. 

Despite the large number of studies on the use of QUS to distinguish between benign and malignant tissue, little research has been done on what fundamental differences there may be between the ultrasonic characteristics of cancerous cells and benign cells.  A few studies have explored if the differences in tissue characteristics extends to the cells themselves. Doyle et al. \cite{Doyle2010} showed differences in the spectral response of healthy compared to malignant breast epithelial cells. There have been other studies on the backscatter coefficients of Chinese hamster ovary (CHO) cells\cite{Teisseire2010}$^{,}$ \cite{Han2011}  including how QUS properties (SOS, attenuation, spectra)  of CHO cells change during cell death. 
In this paper we measured the attenuation and SOS values between 2 and 18 MHz of healthy epithelial prostate cells and malignant epithelial prostate cells, which furthers knowledge about using QUS to distinguish characteristics of benign and malignant prostate tissue.

\section{\label{sec:Meth} Methods}
To measure SOS and attenuation a through transmission narrow band measurement was used. This method has been thoroughly tested and verified using tissue mimicking phantoms as documented by Madsen et al. \cite{Madsen1986} Frequency paired single element unfocused immersion transducers (Olympus A3--S SU, series) 
are aligned with each other in the water tank. A waveform generator (Agilent 33500B) is used to send a single frequency sine pulse (about 20 wavelengths long with a pulse repetition frequency of 500 Hz) at integer frequency values over the bandwidth of the transducers. The pulse is transmitted via one transducer and received by the second, which reads out to an oscilloscope (Tectronix TDS 3014C).  The output of the receiving transducer is averaged on the oscilloscope 512 times. Three sets of paired transducers were used with peak frequencies at 5, 10, and 15 MHz. The peak to peak amplitude and the time of arrival of the output signal is measured for a water only path and for the same path but with the cell pellet inserted. Degassed pure water (filtered by Milli-Q system at 18 M$\Omega$) was used. The water was maintained at the temperature of 31 $^{\circ}$C. 

To find the SOS, the time for sound pulse transit with and without a sample between the transducers is measured. The speed of sound in the water path is well  known\cite{Marczak1997}, and the difference between the two times is used to calculate the speed of sound in the sample. Knowing the width of the sample, measured using digital calipers, the speed of sound is found by:
\begin{equation}\label{eqn:SOS}
c_s = \frac{d c_w}{d-\Delta t c_w},
\end{equation} where $c_s$ is the speed of sound of the sample, $c_w$ is the speed of sound of water, d is the width of the sample, and $\Delta$t is the difference in time between the water only path and the path with the sample inserted. 

Similarly to find the attenuation, the amplitude of the water only path is compared to the amplitude of the path with the sample inserted. Since the cell pellet samples have saran windows a correction for the transmission coefficient of saran is included. The transmission coefficient is given by:  \begin{equation}\label{eqn:TransCoeff}
T^2 =  \frac{4Z_1 Z_3}{(Z_1+Z_3)^2 cos^2(k_2 l)+(Z_2 + \frac{Z_1 Z_3}{Z_2})^2 sin^2(k_2 l)}.
\end{equation}
In this equation $k_2$ is the wavenumber of the sine pulse, the thickness of the saran layer is $l$, $Z_2$ is the acoustic impedance of the saran, $Z_1$ the acoustic impedance of water, and $Z_3$ the acoustic impedance of the sample material. (The acoustic impedance of the sample is found from the SOS measurement and a density measurement.) Using these equations and solving the attenuation coefficient in dB/cm give the general equation: 
\begin{equation} \label{eqn:atten}
\alpha = \frac{20}{d}  \rm{log}_{10}(\frac{A_w}{A_s}  T^2 ),
\end{equation}
where $A_w$ is the amplitude measured for the water path and $A_s$ is the amplitude measured with the sample in the path. 

\begin{figure}[t]
\includegraphics[width=1\textwidth]{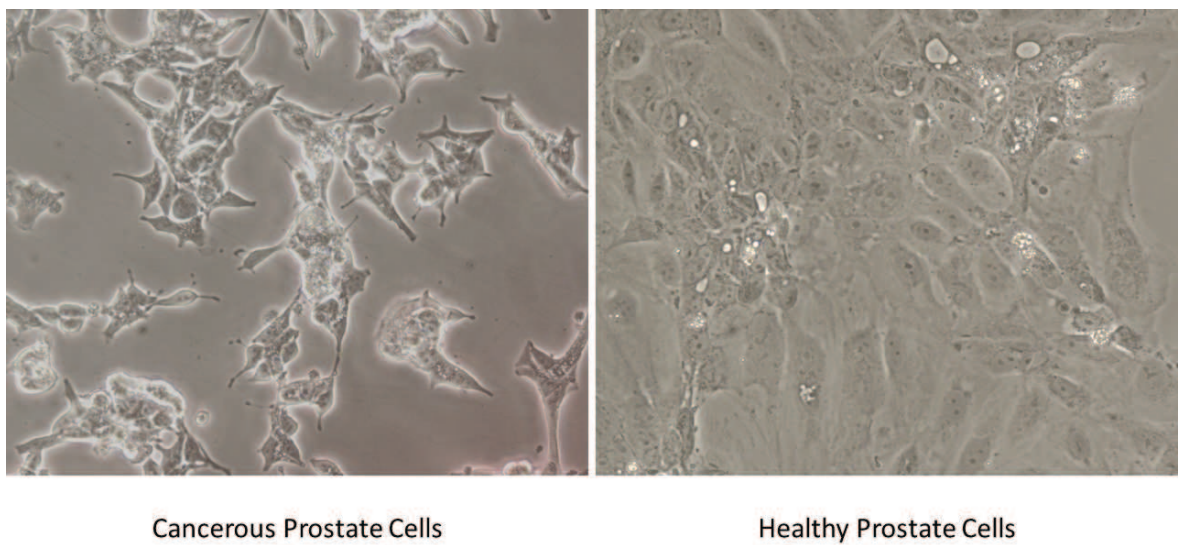}
\caption{\label{Figure1} (Color online) Images of the cell lines; cancerous prostate cells on the left and benign prostate cells on the right. }
\end{figure}

\begin{figure}[t]
\includegraphics[width=1\textwidth]{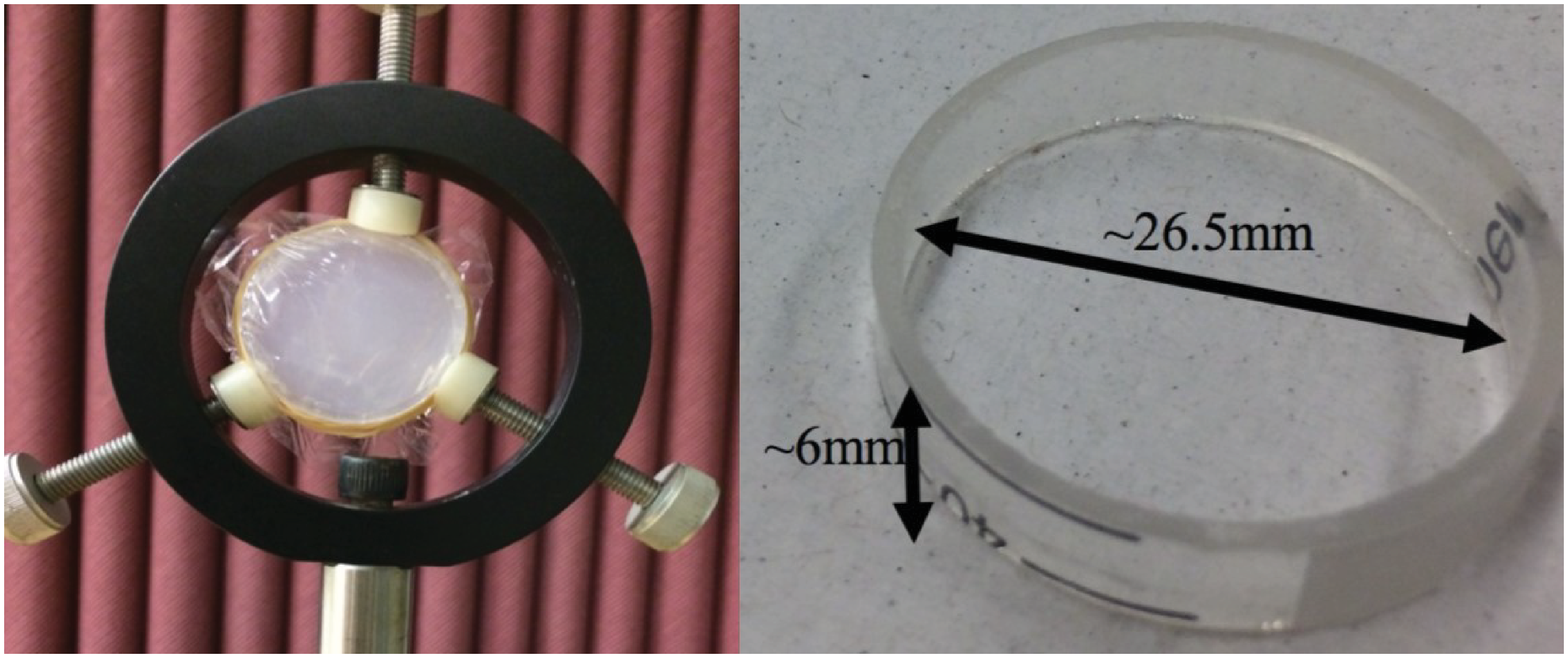}
\caption{\label{Figure2} (Color online) (Left) Image of a completed cell pellet. (Right) Approximate dimensions of the cell pellet container. }
\end{figure}

Measurements were made on two cell lines (1) healthy human prostate epithelial cells (HPC) and (2) malignant human prostate epithelial cells (MPC). Figure \ref{Figure1} shows example pictures of each cell line. 
The cells were cultured, and when a sufficient number of cells of a particular line grew, (one or two 25 $\rm{cm}^2$ flasks with the cells 80\% confluent) a cell pellet was made. The cell pellet holds cells in a form that allows for a reliable scan using US. When the cells are 80\% confluent in a flask, they are detached and a cell count is done. The cell count is used to determine how many cells will be in the pellet (and to control how many cells go into the pellet by finding the cell density). This study aimed for a cell density of about 5 million cells/mL. Next the cells are centrifuged to separate the cells from the supernatant. The cells are suspended in an 1\% agar base, creating a cell pellet.  The pellet is created in a plastic container with saran windows on top and bottom, and is ready to be scanned in the US water tank.  Figure \ref{Figure2} shows a picture of a completed cell pellet. Cell pellets are examined for bubbles or other defects before and after the experiment, and data is only kept for cell pellets that show no significant defects. 
\section{\label{sec:Res}Results}
Speed of sound (SOS) and attenuation were measured for 3 paired sets of unfocused transducers. Measurements were made at 2, 3, 4, 5, 6, 7, and 8 MHz for the transducers with a 5 MHz peak frequency, at 7, 8, 9, 10, 11, 12 and 13 MHz for the transducers with a 10 MHz peak frequency, and at 12, 13, 14, 15, 16, 17, and 18 MHz for the transducers with a 15 MHz peak frequency.  Overlapping frequency measurements ensure the stability of the cell pellet over the experiment as well as the consistency of the results.  A total of 5 cell pellets for the healthy prostate cells (HPCs) and a total of 4 cell pellets for the malignant prostate cells (MPCs) were measured.  Results for the average SOS and average frequency independent attenuation  are reported in Table 1. Table 1 also gives the average density for each type of (HPC or MPC) cell pellet. For every cell pellet 2 to 6 regions in the hemocytometer were counted and the counts averaged together.
Error for the individual cell pellet density was determined by using the standard deviation between counts from the hemocytometer. Reported densities in Table 1 are the average of all cell pellets for each type. The error in the average is the standard deviation between the densities of all cell pellets of that type, which dominates the error from the individual counts. 

Error for the SOS measurements was determined for each individual cell pellet by the standard deviation over three separate time measurements and three separate width measurements. The error of SOS in water was minimal as the temperature was stable throughout the experiment, with redundant measurements using a digital thermometer and an alcohol thermometer at two different locations in the water tank.  The error in width and time was propagated through the equation for finding the SOS in the sample, equation \ref{eqn:SOS}, giving an error for each measurement at each frequency.  The values for SOS were averaged at each frequency over all the measured cell pellets for each cell line. For the HPCs, the SOS measurements spanned a total of 7 m/s, while for the MPCs the measurements spanned a total of 5 m/s. No significant dispersion with increasing frequency was observed. The final reported values is an average over frequency for each cell line. The standard error over all measurements is given and dominated the propagated error for each individual measurement.
\begin{figure}[t]
\includegraphics[width= 1\textwidth]{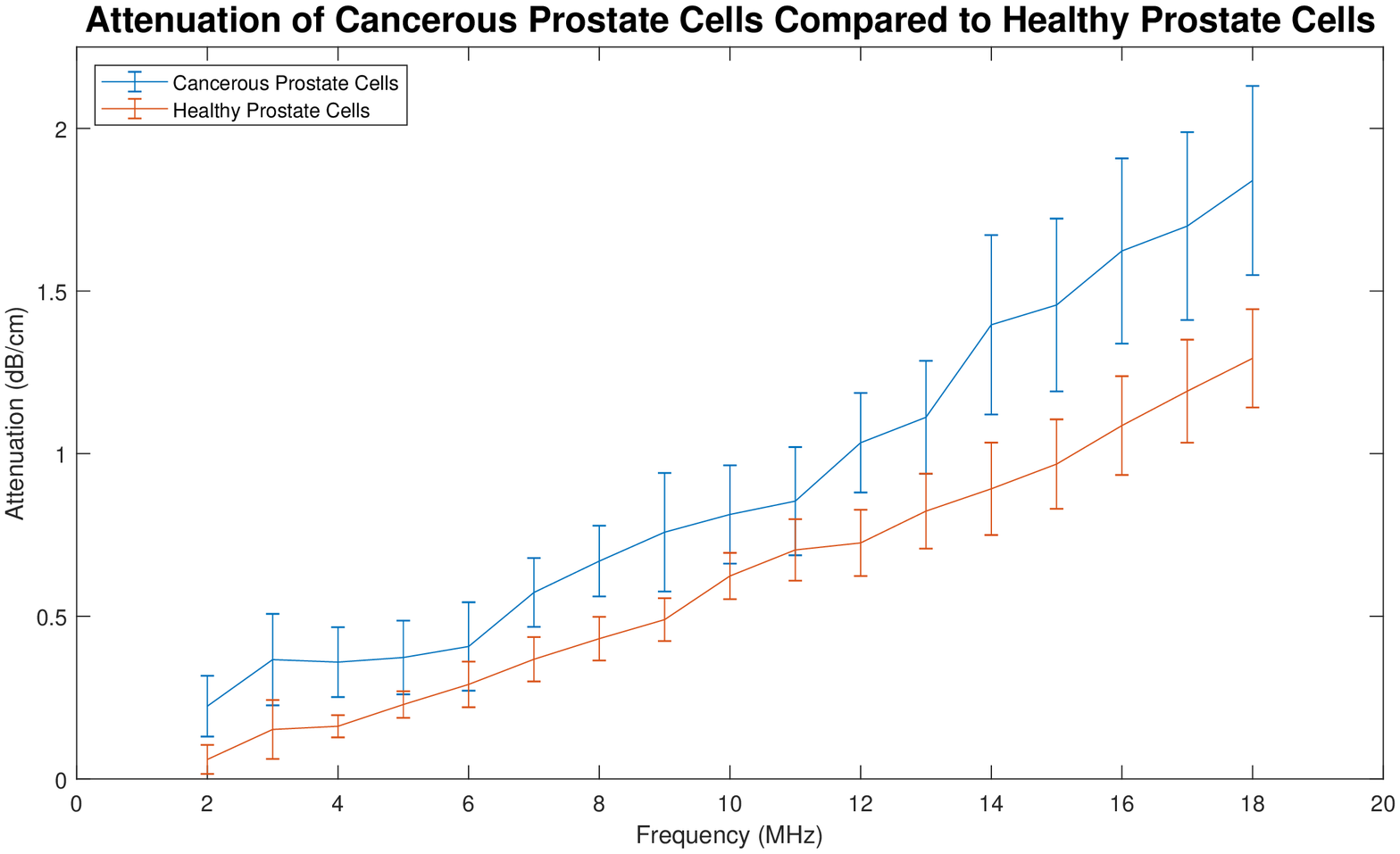}
\caption{\label{Figure3} (Color online) Attenuation as a function of frequency for healthy prostate cells (red) and cancerous prostate cells (blue). Error bars are given for each data point. Cancerous prostate cells show a consistent higher attenuation. }
\end{figure}

Figure \ref{Figure3} shows the attenuation as a function of frequency for both the MPCs and HPCs. The error in the attenuation values for each cell pellet and frequency are found using the standard deviation in the three repeated measurements of width and standard deviation for the three repeated amplitude measurements for both water and sample paths.  These errors are propagated through the equation for attenuation, equation \ref{eqn:atten}. Error for the transmission coefficient of saran and value of the water correction were not included since they were at least 2 orders of magnitude smaller and did not contribute significantly.  For each cell pellet the attenuation values were averaged over the same frequency, and the error added as the root of the sum of squares. The standard error due to multiple measurements was also calculated. The final total error is the root of the sum of the squares of the standard error and the propagated error values. 
\begin{table}[ht]
\caption{\label{tab:table1} For each cell line density, attenuation, and speed of sound averages are given with uncertainties}
\begin{tabular}{cccccccc}
  & $\rho$\footnotemark[1] & SD\footnotemark[2] & SOS \footnotemark[3]
 & Un. \footnotemark[4] & $\alpha$ \footnotemark[5]& Un. \footnotemark[6]\\
   &  (\#/mL) &(\#/mL) & (m/s)
 & (m/s) & $\frac{\rm{dB}}{\rm{cm-MHz}}$ & $\frac{\rm{dB}}{\rm{cm-MHz}}$\\
\hline
HPC & 4.78 x $10^6$ & 8.1 x $10^5$ & 1521.5 & 0.6 & 0.057 & 0.003 \\
MPC & 4.8 x $10^6$ & 1.6 x $10^6$ & 1521.4 & 0.8 & 0.091 & 0.003 \\

\end{tabular}
\footnotetext[1]{Cell Density}
\footnotetext[2]{Standard Deviation in Cell Density}
\footnotetext[3]{Speed of Sound}
\footnotetext[4]{Uncertainty in Speed of Sound}
\footnotetext[5]{Average Attenuation Coefficient  for all Frequencies}
\footnotetext[5]{Uncertainty in Average Attenuation Coefficient}
\end{table}
The frequency independent attenuation is also reported in Table 1. These values are the average of the attenuation divided by frequency values. Although attenuation as a function of frequency is not linear, the power relationship is small enough that the average value is reasonably representative for each cell line. The error in these averages is given by the standard error of the calculations. 

\section{\label{sec:Con}Conclusion}
The quantitative ultrasonic properties of speed of sound and attenuation were measured for two cell lines, a malignant prostate cell line and normal epithelial prostate cells, from 2 to 18 MHz. Both cell lines have equivalent values for speed of sound. The cancerous cell line has statistically significant higher values for attenuation, showing fundamental differences in interaction with US for benign and malignant prostate cells. 

Further studies on these two cell lines should explore the scattering properties of the cells as a function of frequency, as this has also been shown as a method for distinguishing between healthy and malignant prostate tissue. 
For the experiments in this paper the cells were individually suspended in agar, while other papers have made measurements on cells in a mono layer. \cite{Doyle2010} Comparison of measurements for suspended cells  verses mono layers may tell us more about the QUS characteristics of cells and how those ultrasonic properties depend on cell distribution.


\end{document}